\documentclass[12pt,preprint]{aastex}

\slugcomment{Accepted for publication, Astrophysical Journal Letters}

\shorttitle{The first early-type L subdwarf}
\shortauthors{Lepine, Rich, \& Shara}

\begin{document}

\title{LSR1610-0040: the first early-type L subdwarf}

\author{S\'ebastien L\'epine\altaffilmark{1,2,3}, R. Michael
Rich\altaffilmark{4}, and Michael M. Shara\altaffilmark{1}}
\altaffiltext{1}{Department of Astrophysics, Division of Physical
Sciences, American Museum of Natural History, Central Park West at
79th Street, New York, NY 10024, USA, lepine@amnh.org, shara@amnh.org}
\altaffiltext{2}{Kalbfleich research fellow}
\altaffiltext{3}{Visiting Astronomer, MDM observatory}
\altaffiltext{4}{Department of Physics and Astronomy, University of
California at Los Angeles, Los Angeles, CA 90095, USA,
rmr@astro.ucla.edu}

\begin{abstract}
We report the discovery of LSR1610-0040, a previously unreported
$r=17.5$ star with a very high proper motion $\mu=1.46\arcsec$
yr$^{-1}$. This very cool star ($b-i$=6.3) is found to have a peculiar
spectrum that does not fit into the standard sequence of late-M dwarfs
and L dwarfs. Rather, the spectrum is more typical of an ultra-cool
subdwarf, with weak bands of TiO and no detectable VO. But because
LSR1610-0040 is so much redder than any other sdM known, and because
it does not appear to fit well into the sdM sequence, we propose that
LSR1610-0040 be assigned a spectral type sdL, making it the first
early-type L subdwarf known. Evidence suggests that LSR1610-0040 is an
old, metal-poor star whose mass is just above the limit of hydrogen
burning ($M\approx0.08$M$_{\sun}$).
\end{abstract}


\section{Introduction}

Recent years have seen the discovery, through the use of large
infrared surveys, of numerous objects extending the main sequence
down to, and beyond the limit of hydrogen burning
\citep{KRLCNBDMGS99}, and well into the substellar regime
\citep{BKBRBLMGDMCS02}. This sequence of objects, which go from
spectral type M, to spectral type L, to spectral type T, consists
almost exclusively of relatively metal-rich, red dwarfs and brown
dwarfs. To this date, however, no equivalent spectroscopic sequence
exists for subdwarfs. Cool, red, M subdwarfs (sdM) are the metal-poor
equivalent of the red dwarfs, and they are the low-mass
representatives of Population II stars. The current spectroscopic
sequence of M subdwarfs does not extend beyond spectral subtype
sdM8.0 \citep{LSR03}, and very few ``ultra-cool subdwarfs'' (spectral
subtype sdM6.0 and later) are known. As a result, the physical
properties of old, very low mass stars in the Galaxy are poorly
constrained.

Nevertheless, extremely cool subdwarf stars of spectral type later
than sdM8 are likely to exist in small but significant numbers. Deep
photometry of the globular cluster Messier 4 strongly suggest that its
main sequence extends beyond spectral type sdM8
\citep{RBFGHIKLRSSS02}. If very cool, hydrogen burning ``L subdwarfs''
(sdL) do exist in globular clusters, then they are most likely to
exist in the thick disk or halo as well and should be detected in the
field. They should however be relatively rare, as they correspond to a
thin slice of the mass range just above the H-burning limit. Estimates
place them at less than 1 in 30,000 of stars in the solar
neighborhood. But their probability of detection should be
significantly increased in samples of high proper motion stars, which
are strongly biased towards high-velocity (halo) objects. Nearby L
subdwarfs are thus most likely to be identified among large samples of
faint, high proper motion stars.

Recently, \citet{Betal03} have identified what is most likely the
first metal-poor L subdwarf. This candidate sdL, known as 2MASS
0532+8246, is a faint object with a large proper motion
($\mu\simeq2.6\arcsec$ yr$^{-1}$) identified in the 2MASS survey from
its large optical to infrared color (it has no detectable counterpart
in the in the blue and red Palomar Sky Survey plates). While its
spectral energy distribution is analogous to an L7 dwarf, 2MASS
0532+8246 shows unusually strong lines of TiO and metal hydrides in
the optical, as well as an unusually strong collision-induced H$_2$
absorption in the infrared. Evidence suggest it is substellar. This
discovery opens the prospect for obtaining a classification
sequence for metal-poor objects down to the hydrogen-burning limit and
beyond. At this point, there remains a wide gap to be filled between
the known ultra-cool sdM and 2MASS 0532+8246, before a consistent
sequence can be established.

In this paper, we report the discovery of LSR1610-0040, a very red
star with a large proper motion $\mu=1.46\arcsec$ yr$^{-1}$. Our
spectroscopy reveals that the star shows features associated with
ultra-cool M subdwarfs, although it appears to be significantly
cooler than the coolest sdM known to date, the sdM8.0 star
LSR1425+7102. As it appears to be intermediate between ultra-cool sdM
stars and the candidate sdL object 2MASS 0532+8246, we propose
that LSR1610-0040 is the first known example of an ``early-type'' L
subdwarf.

\section{Proper Motion Discovery and Photometry}

The high proper motion star LSR1610-0040 was discovered as part of
our new search for high proper motion stars in the northern sky using
the Digitized Sky Survey \citep{LSR02}, performed as a part of the
NStars initiative. The Digitized Sky Survey discovery fields are
presented in Figure 1. The left panel shows the 1953 POSS-I red
(xx103aE + plexi) image of a $4.25\arcmin\times4.25\arcmin$ field
centered on the position of the star at epoch 2000.0. The right hand
side shows the 1992 POSS-II red (IIIaf + RG 610) image of the same
field. The large temporal baseline makes the motion of LSR1610-0040
very obvious; the star is moving at a rate
$\mu_{RA}=-0.77\arcsec$yr$^{-1}$, $\mu_{DEC}=-1.23\arcsec$yr$^{-1}$.

The star is not recorded in the LHS catalog \citep{L79} and a search
on Simbad (http://simbad.u-strasbg.fr/Simbad) around that location on
the sky yielded no results, confirming that this high proper motion
star is being reported here for the first time. We have found
counterparts of the star in the USNO-B1.0 catalog \citep{Metal03}.
The object is identified from the blue, red and infrared POSS-II plates
and the calculated USNO-B1.0 photographic magnitudes are $b=21.1$,
$r=17.5$, and $i=14.8$. The USNO-B1.0 lists this star as having a
proper motion [$\mu$RA,$\mu$DE]=[+162,-370] mas yr$^{-1}$, which
suggests an incorrect pairing to a first epoch object on the POSS-I
plates. Indeed, the USNO-B1.0 gives for that star a first epoch POSS-I
blue magnitude $b_1=21.2$ but no first epoch POSS-I red magnitude
($r_1=0.0$). As it turns out, visual examination of scanned survey
images reveal that the star does not show up on the POSS-I blue
(103aO) plate at the position extrapolated from the proper motion,
although is shows up very clearly on the POSS-I red (103aE) plate.
But the reality of this high proper motion star cannot be cast into
doubt, as the star clearly shows up on 10 other optical survey plates:
as a faint star on 2 different SERC-EJ (IIIaJ) plates, as a much
brighter star on 2 other SERC-ER (IIIaF), and also on 2 each of
POSS-II blue (IIIaJ), red (IIIaF), and infrared (IVN) plates. In each
case the position accurately matches the extrapolated proper motion of
the star at the epoch of the plate.

The high proper motion star also shows up as a bright star in the
2MASS survey, and we find it to be a perfect match to a bright
object in the 2MASS All-Sky Point Source Catalog. Identified as 2MASS
J16102900-0040530, it has infrared magnitudes $J=12.91$, $H=12.32$,
and $K_s=12.02$ (see Table 1).

\section{Spectroscopy}

The star LSR1610-0040 was observed on the night of 19 February 2003, at
the 2.4m Hiltner telescope of the MDM Observatory. A spectrum of the
star was obtained with the MkIII spectrograph equipped with
2048$\times$2048 front-side illuminated, thick LORAL CCD
(``Wilbur''). We used a 300 l/mm grating blazed at 7500\AA, with a red
order blocking filter. The star was imaged through a 0.8$\arcsec$ slit,
yielding a nominal spectral resolution of 7\AA. Standard spectral
reduction was performed with IRAF using the CCDPROC and SPECRED
packages, including removal of telluric features. Calibration was
derived from observations of the standards Feige 66 and Feige 67
\citep{MG90}. Both the target and the standard were observed at the
smallest possible airmass ($<1.3$) and with the slit at the
parallactic angle to minimize slit loss due to atmospheric
diffraction, providing excellent spectrophotometric calibration.

The spectrum of the star is displayed in Figure 2. The main spectral
features of the very red object are identified. The most prominent
features are the molecular bands of CaH, TiO, FeH, CrH, and H$_2$O,
and atomic lines of \ion{K}{1}, \ion{Rb}{1}, and \ion{Na}{1}. We
measured the radial velocity of the star from the centroids of the
\ion{K}{1} $\lambda\lambda$7665,7699 and \ion{Rb}{1}
$\lambda\lambda$7800,7947 atomic absorption lines. After correction
for the earth's motion in space, and accounting for uncertainties, we
find that the star has a large heliocentric radial velocity
V$_{hel}=-130\pm15$ km s$^{-1}$.

%%% LINES
%
%   KI-7664.911  => 7661.1 , EQW>20   , -149
%   KI-7698.974  => 7695.1 , EQW>20   , -151
%  RbI-7800.259  => 7796.0 , EQW=3.8  , -163
%    ?           => 7907.9 , EQW=0.8  ,
%  RbI-7947.597  => 7943.3 , EQW=3.4  , -162
%  CsI-8521
%  CsI-8943
%  NaI-8183.255  =>(8179.3)            (-144)
%  NaI-8194.824  =>(8190.3)            (-165)
%                                       ===
%                                      -156 + 28
%                             V_helio=[-130\pm15km/s]

\section{Spectral classification}

A comparison with the standard sequence of M dwarfs \citep{KHM91} and
L dwarfs \citep{KRLCNBDMGS99}, shows no clear similarity between
LSR1610-0040 and any single one of the subtypes of the metal-rich
sequence. Rather, our star shows a variety of features which are
reminiscent of both late-type M dwarfs and early to mid-type L
dwarfs. While the spectral energy distribution of LSR1610-0040 is
comparable to that of a late-type M dwarf, the molecular bands of TiO
beyond 7500\AA\ are unusually weak. The VO bands, which are strong in
all late-type M dwarfs and early-type L dwarfs, are completely absent
(or extremely weak) in LSR1610-0040. The spectral features most at
odds with an M dwarf classification are the two prominent atomic lines
of \ion{Rb}{1}, which are normally seen only in L dwarfs. The
detection of the FeH and CrH bands redwards of 8000\AA\, especially
the relatively strong FeH bandhead at $\lambda9896$ are also
reminiscent of early-L to mid-L spectra, although the bands do appear
to be relatively weaker. On the other hand, the strong CaH and TiO
bands around 7000\AA\ are completely at odds with the standard L dwarf
sequence.

On the other hand, the spectrum of LSR1610-0040 bears a significant
resemblance to the spectrum of the ultra-cool sdM8.0 star
LSR1425+7102 \citep{LSR03}. In particular, both stars have strong CaH
$\lambda6750$ and TiO $\lambda7053$ bands, but very weak TiO
$\lambda7589$ and $\lambda8432$ as weel as weak CrH and FeH. The  main
difference are the RbI lines, absent in LSR1425+7102, and the
significantly redder color of LSR1610-0040. Overall, it looks like
the spectrum of LSR1610-0040 would be consistent with the star being a
new type of extremely cool subdwarf. Thus, the many peculiarities in
the spectrum of LSR1610-0040, when it is compared to the M-L dwarf
sequence, could be explained simply in terms of a low metal abundance.

But LSR1610-0040 does not fit well into the standard sdM
sequence. Spectral subtypes for sdM stars are based on the
strengths of the CaH and TiO molecular bands around 7000\AA\
\citep{RHG95}, as measured by the CaH2, CaH3, and TiO5 indices. For
LSR1610-0040 we measure CaH2=0.257, CaH3=0.478, and
TiO5=0.295. Applying these values to the standard quantitative
classification scheme of M subdwarfs, developed by \citet{G97} and
expanded by \citet{LSR03}, nominally yields a spectral type
sdM6.0. However, this spectral assignment is completely inconsistent
with the spectral energy distribution of LSR1610-0040. To quantify
this, we use the Color-M index, which measures the slope of the
spectrum in the 6500\AA-8000\AA\ range \citep{LRS03}. The value we find
for LSR1610-0040 (Color-M=4.7) is significantly larger than the
typical value for sdM6.0 stars (Color-M$\simeq$2.2), even larger than
the value we measured in the only known sdM8.0 star LSR1425+7102
(Color-M=2.4). The extreme color of LSR1610-0040 is also confirmed by
the optical photometry: with a $b-i$=6.3, the star is significantly
redder than LSR1425+7102 ($b-i$=4.6). The same holds true for the
infrared to optical color: LSR1610-0040 has $r-K_s=5.5$, while the
sdM8.0 has $r-K_s=4.3$.

This brings the possibility that LSR1610-0040 should be
considered an ``L subdwarf'' (sdL), by extension of the M-L dwarf
sequence to an equivalent sdM-sdL sequence. Such a sequence
is, however, not yet defined beyond spectral type sdM8.0, and the only
other known candidate sdL to this date is the extremely cool
metal-poor object 2MASS 0532+8246 \citep{Betal03}. Spectroscopic
features in 2MASS 0532+8246 which make it stand apart from the L dwarf
sequence could help explain some of the ``anomalies'' observed in
LSR1610-0040. In particular, the TiO bands are unusually strong in 2MASS
0532+8246. If this is true of the whole sdL sequence, then the strong
TiO $\lambda7053$ band observed in LSR1610-0040 might no be so unusual
for an early-type sdL. But overall, the fact that 2MASS 0532+8246 is
evidently so much cooler than LSR1610-0040 makes any comparison
between the two stars difficult.

In order to shed further light on the status of LSR1610-0040, we
compare its spectral energy distribution to the theoretical models of
low mass, metal-poor stars computed by \citet{BCAH97}. Assuming that
LSR1610-0040 is related to M subdwarfs, which have a
metallicity around $[m/H]\sim-1.2$ \citep{G97}, we compare it to
models with $[m/H]=-1.0$, $-1.3$, and $-1.5$. We conclude that
LSR1610-0040 is most consistent with a hydrogen burning object with a
mass no larger than 0.085M$_{\odot}$, and with an effective
temperature in the range 2100K-2500K. Exact values for the mass and
effective temperatures are dependent on the metallicity of the object;
a detailed spectroscopic modeling will thus be required to pinpoint
the fundamental parameters of the star. In any case, the estimated
range in effective temperature (2100K-2500K) is consistent with the
effective temperatures generally assigned to metal-rich early-type L
dwarfs \citep{D02}.

Hence we have a cool star whose spectrum clearly suggests a metal-poor
composition, whose color is much redder than any other known sdM
stars, and for which the standard spectral classification scheme of
\citet{G97} for sdM stars is invalid. Furthermore, its colors are
consistent with a hydrogen burning star just above the hydrogen
burning limit and with an effective temperature in the range normally
assigned to early-type L dwarfs. We therefore propose that
LSR1610-0040 should be considered the first prototype of an {\em
early-type L subdwarf}, by opposition to the much cooler sdL object
2MASS 0532+8246 found by \citet{Betal03}, which may now be considered
a {\em late-type L subdwarf}.

\section{Distance and kinematics}

The general spectral energy distribution of LSR1610-0040 is similar to
that of a late-type M dwarf. The optical to infrared color
$r-K_s\simeq5.5$ is actually similar to that of an M6 dwarf
\citep{KRLCNBDMGS99}; the infrared colors $J-H=0.59$ and $H-Ks=0.30$
are also typical of a late-type M dwarf. The absolute $J$ magnitude of
an M6 dwarf is $M_J\simeq10.5$ \citep{D02}. With an $M_J=10.5$,
LSR1610-0040 would be at a distance of 30pc. This is clearly an
absolute, upper limit, because subdwarfs are all underluminous
relative to dwarfs of the same color. If we instead use the
observation that subdwarfs of a given {\em spectral subtype} tend to
have a similar absolute magnitude {\em in the infrared} as dwarfs of
the same spectral subtype \citep{LRS03}, than we may assume that
LSR1610-0040 has the luminosity of an early-type L dwarf, with
$M_J\simeq12$ \citep{D02}. Under this assumption, LSR1610-0040 would
have a distance modulus of about 1.0, placing it at about 15 parsecs
from the Sun. Using the absolute magnitude calibrations of metal-rich
stars for LSR1610-0040 is, however, a risky proposition at this
point.

Alternatively, we can estimate a distance for LSR1610-0040 based on
the models of \citet{BCAH97}. A 0.083M$_{\sun}$ star with a metallicity
$[M/H]$=-1.0 to -1.5 should have absolute magnitudes $M_R\approx16.3$,
$M_I\approx13.8$, and $M_J\approx11.8$. These values suggest a
distance modulus $\approx1.1$ for LSR1610-0040, which is consistent
with the estimate given above. Given all the uncertainties, we conclude that
a distance modulus in the range 0.5-1.5 is a reasonable estimate for
LSR1610-0040, and we thus conservatively place the star at a distance of
12-20pc, or d=$16\pm4$pc

%Which gives it: R-Ks~5.5  and J-Ks=0.9 not at all like 
%any L dwarf, more like an M6 dwarf. With J-H=0.59
%and H-Ks=0.30 it also fall in the loci of M dwarfs in
%the J-H/H-K diagram.
%
%If it wasn't for the weird spectrum, I'd say it's an M6.
%But if it was M6, it would be at about 30pc, and the
%proper motion would yield a transverse velocity
%V=200 km/s, not like your typical  disc dwarf!
%
%On the other hand, if it has the luminosity of an 
%early L dwarf (absolute J~12), then it is at about 
%10-15 parsecs.

Based on this distance, the large proper motion yields an estimated
transverse velocity $V_t=110\pm30$km s$^{-1}$. Combined with the radial
velocity measurement, we calculate a space motion $[U,V,W] =
[-117\pm18,-108\pm24,-10\pm19]$ km s$^{-1}$ relative to the local
standard of rest, where $U$ is towards the Galactic center, $V$
towards the direction of Galactic rotation, and $W$ towards the north
Galactic pole. This places LSR1610-0040 just outside the $2\sigma$
limits of disk stars, as defined by \citet{CB00}. The relatively low
value of $W$ makes it most likely for LSR1610-0040 to be an
Intermediate Population II star, i.e. a possible member of the old
disk, although a halo membership cannot be excluded. In any case, the
kinematics supports the idea that LSR1610-0040 is an old star.

\section{Conclusions}

We have discovered a cool star with a large proper motion
$\mu=1.46\arcsec$ yr$^{-1}$. Spectroscopy reveals that the star
doesn't fit in the standard M dwarf and L dwarf sequence, and that it
is most likely a metal-poor object. The star is shown to be much
redder than any other known subdwarf, and cannot be fitted in the
sequence of M subdwarfs. The star appear to be somewhat intermediate
between the ultra-cool sdM8.0 star LSR1425+7102 and the recently
discovered L subdwarf 2MASS 0532+8246. We suggest that
LSR1610-0040 is the first example of an ``early-type'' L subdwarf.

This star should be considered a high priority target for astrometric
parallax determination. Because it is a relatively bright object in
the red (i=14.8) and infrared (J=12.9), LSR1610-0040 is an easy target
for follow-up observations. Absolute magnitudes will help determine
whether this star stands at the very limit of hydrogen burning for a
metal poor star. Available data suggest that LSR1610-0040 is most
likely a $\approx$0.08M$_{\sun}$ subdwarf at a distance
$d=16\pm4$pc. Detailed spectroscopic modeling will be required to
determine the abundance and surface gravity of the object.

\acknowledgments

S.L. is a Kalbfleich research fellow of the American Museum of Natural
History. This research program is being supported by NSF grant
AST-0087313 at the American Museum of Natural History, as part of the
NStars initiative.

\newpage

\begin{deluxetable}{lrl}
\tabletypesize{\scriptsize}
\tablecolumns{3} 
\tablewidth{0pt} 
\tablecaption{Basic Data for LSR1610-0040} 
\tablehead{Datum & Value & Units}
\startdata 
RA (2000.0) & 16 10 28.85 & h:m:s\\
DEC (2000.0)& -00 40 53.0 & d:m:s\\
$\mu$       & 1.46 & $\arcsec$ yr$^{-1}$\\
pma         & 212.0 & $\degr$\\
v$_{hel}$   & -130  & km s$^{-1}$\\
b\tablenotemark{1}     &   21.1 $\pm$0.3  & mag\\
r                      &   17.5 $\pm$0.3  & mag\\
i                      &   14.8 $\pm$0.3  & mag\\
J\tablenotemark{2}     &   12.91$\pm$0.02 & mag\\
H                      &   12.32$\pm$0.02 & mag\\
K$_s$                  &   12.02$\pm$0.03 & mag\\ 
Spectral Type & sdL:& \\
Distance & 16$\pm$4 & pc \\
$U$ &    -111$\pm$18 & km s$^{-1}$\\
$V$ &     -85$\pm$29 & km s$^{-1}$\\
$W$ &     -24$\pm$21 & km s$^{-1}$
\enddata
\tablenotetext{1}{Photographic B, R, and I magnitudes from USNO B-1.0 catalog.}
\tablenotetext{2}{Infrared J, H, and K$_s$ magnitudes from 2MASS
All-Sky Point Source Catalog.}
\end{deluxetable} 

\newpage

\begin{figure}%1
\plotone{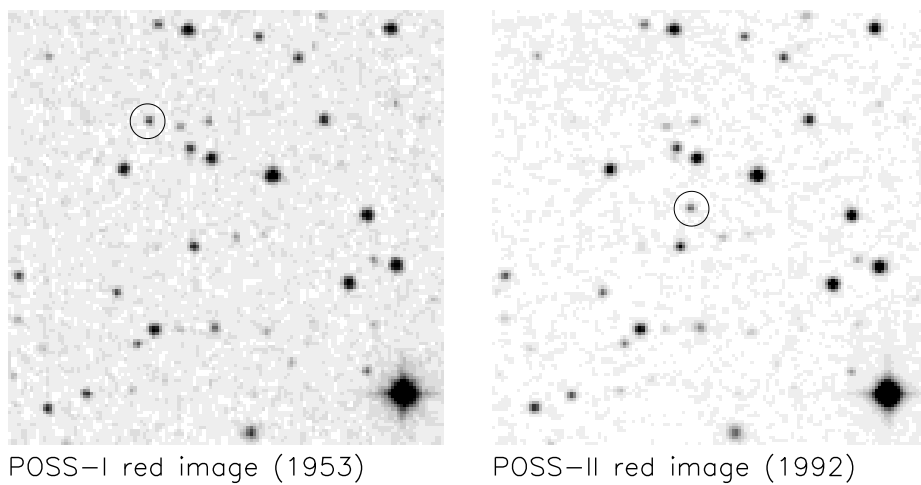}
\caption{\label{fig1} The new high proper motion star
LSR1610-0040. Left: red plate of the first epoch Palomar Sky Survey,
obtained in 1953. Right: red plate of the second epoch Palomar Sky
Survey, obtained in 1992. Both fields are $4.25\arcmin$ on the side,
with north up and east left. Circles are drawn centered on the
location of LSR1610-0040 at each epoch. The star is moving with a
proper motion $\mu=1.46\arcsec$yr$^{-1}$.}
\end{figure}

\begin{figure}%2
\plotone{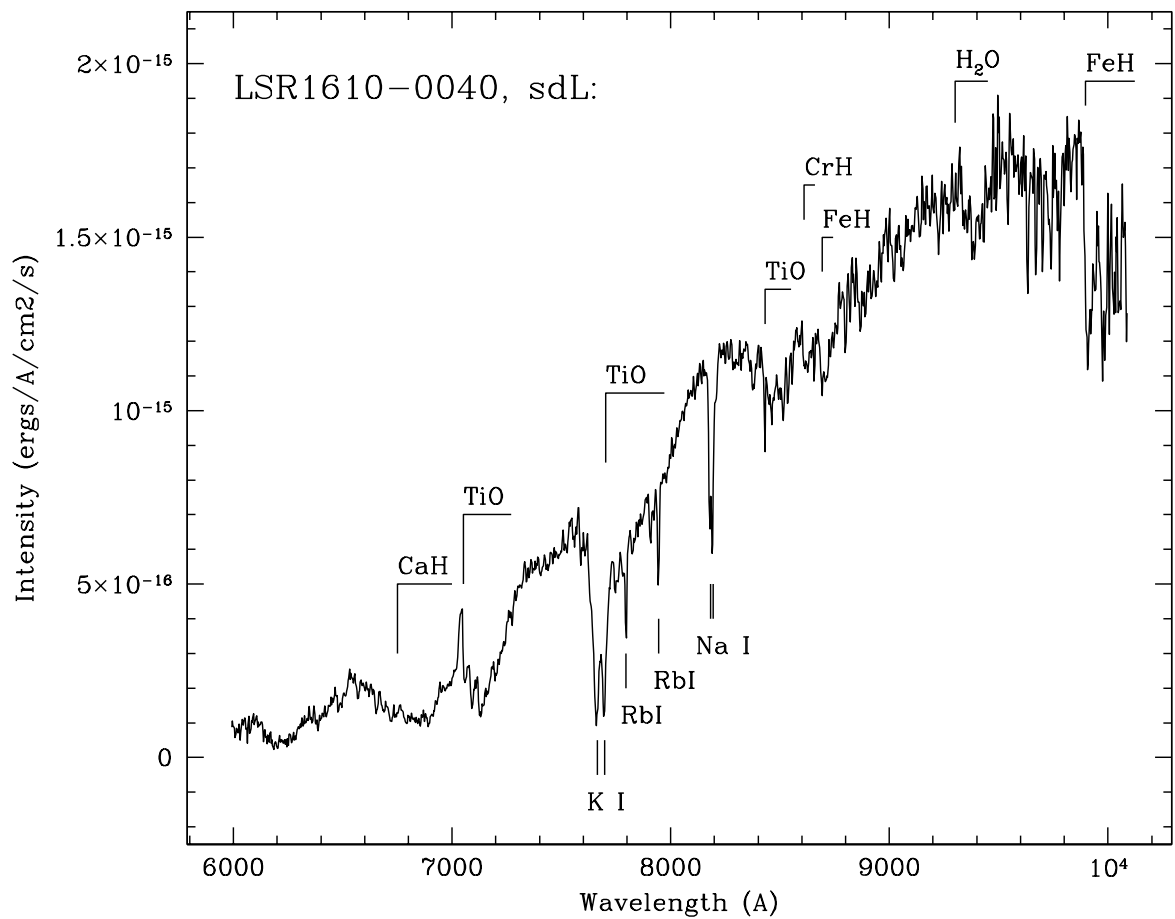}
\caption{\label{fig2} Optical spectrum of the high proper motion star
LSR1610-0040. Note the extreme slope of the pseudo-continuum,
indicative of a low temperature. Also note the weakness of the TiO
bands and the complete absence of the VO bands, usually prominent in
late-type M dwarfs. The strong \ion{Rb}{1} lines are reminiscent of L
dwarfs, although the hydride bands are unusually weak. We interpret
this spectrum as a clear indication that the star is both very cool
and metal-poor, and assign a spectral type ``sdL:''.}
\end{figure}


\begin{thebibliography}{}

\bibitem[Baraffe {\it et al.}(1997)]{BCAH97}
Baraffe, I., Chabrier, G., Allard, F., \& Hauschildt, P. H. 1997,
\aap, 327, 1054

\bibitem[Burgasser {\it et al.}(2002)]{BKBRBLMGDMCS02}
Burgasser, A. J., Kirkpatrick, J. D., Brown, M. E., Reid,
I. N., Burrows, A., Liebert, J., Matthews, K., Gizis, J.
E., Dahn, C. C., Monet, D. G., Cutri, R. M., Skrutskie,
M. F. 2002, \apj, 564, 421

\bibitem[Burgasser {\it et al.}(2003)]{Betal03}
Burgasser, A. J., Kirkpatrick, J. D., Burrows, A., Liebert, J., Reid,
I. N., Gizis, J. E., McGovern, M. R., Prato, L., \& McLean,
I. S. 2003, \apj, {\it in press}

\bibitem[Chiba \& Beers(2000)]{CB00} 
Chiba, M., \& Beers, T. 2000, \aj, 119, 2843

\bibitem[Dahn {\it et al.}(2002)]{D02}
Dahn, C. C., Harris, H. C., Vrba, F. J., Guetter, H. H., Canzian, B.,
Henden, A. A., Levine, S. E., Luginbuhl, C. B., Monet, A. K. B.,
Monet, D. G., Pier, J. R., Stone, R. C., Walker, R. L., Burgasser,
Adam J., Gizis, J. E., Kirkpatrick, J. D., Liebert, J., \& Reid,
I. N. 2002, \aj, 124, 1170

\bibitem[Gizis(1997)]{G97}
Gizis, J. E. 1997, \aj, 113, 806 (G97)

\bibitem[Kirkpatrick, Henry, \& McCarthy(1991)]{KHM91}
Kirkpatrick, J. D., Henry, T. J., \& McCarthy, D. W. 1991, \apjs,
77, 417

\bibitem[Kirkpatrick {\it et al.}(1999)]{KRLCNBDMGS99}
Kirkpatrick, J. D., Reid, I. N., Liebert, J., Cutri, R. M., Nelson,
B., Beichman, C. A., Dahn, C. C., Monet, D. G., Gizis, J. E., \&
Skrutskie, M. F. 1999, \apj, 519, 802

\bibitem[L\'epine, Shara, \& Rich(2002)]{LSR02}
L\'epine, S., Shara, M. M., \& Rich, R. M. 2002, \aj, 124, 1190

\bibitem[L\'epine, Shara, \& Rich(2003)]{LSR03}
L\'epine, S., Shara, M. M., \& Rich, R. M. 2003, \apjl, 585, L69

\bibitem[L\'epine, Rich, \& Shara(2003)]{LRS03}
L\'epine, S., Rich, R. M., \& Shara, M. M. 2003, \aj, 125, 1598

\bibitem[Luyten(1979)]{L79}
Luyten W. J. 1979, LHS Catalogue: a catalogue of stars with proper
motions exceeding 0.5" annually, University of Minnesota,
Minneapolis ({\it CDS-ViZier catalog number I/87B})

\bibitem[Massey \& Gronwall(1990)]{MG90}
Massey, P., \& Gronwall, C. 1990, \apj, 358, 344

\bibitem[Monet {\it et al.}(2003)]{Metal03}
Monet, D. G., {\it et al.} 2003, \aj, 125, 984

\bibitem[Reid, Hawley, \& Gizis(1995)]{RHG95}
Reid, I. N., Hawley, S. L., \& Gizis, J. E. 1995, \aj, 110, 1838

\bibitem[Richer {\it et al.}(2002)]{RBFGHIKLRSSS02}
Richer, H. B., Brewer, J., Fahlman, G. G., Gibson, B. K., Hansen,
B. M., Ibata, R., Kalirai, J. S., Limongi, M., Rich, R. M., Saviane,
I., Shara, M. M., \& Stetson, P. B. 2002, \apj, 574, L151

\end{thebibliography}
\end{document}